\documentclass[preprint,aps,prb,showpacs,preprintnumbers]{revtex4}

\usepackage{amssymb}
\usepackage{amsmath}
\usepackage{dcolumn}
\usepackage{bm}
\usepackage{graphicx}
\usepackage{verbatim}
\usepackage{color}
\usepackage{textcomp}

\begin{document}

\preprint{AIP/123-QED}

\title[Sample title]{Signal-to-noise ratio in dual-gated silicon nano\-ribbon field-effect sensors}

\author{A. Tarasov}
\altaffiliation{These authors contributed equally to this work}
\author{W. Fu}
\altaffiliation{These authors contributed equally to this work}
\author{O. Knopfmacher}
\author{J. Brunner}
\author{M. Calame}
\author{C. Sch{\"o}nenberger}
\email{Christian.Schoenenberger@unibas.ch}

\affiliation{Department of Physics, University of Basel,
Klingel\-berg\-strasse 82, CH-4056 Basel, Switzerland}

\date{December 12, 2010}
\definecolor{darkgreen}{rgb}{.004,.68,.31}
\begin{abstract}
Recent studies on nano\-scale field-effect sensors reveal the crucial importance of the low frequency noise for determining
the ultimate detection limit. In this letter, the $1/f$-type noise of Si nano\-ribbon field-effect sensors is investigated.
We demonstrate that the sig\-nal-to-noise ratio can be increased by almost two orders of magnitude if the nano\-ribbon is
operated in an optimal gate voltage range. In this case, the additional noise contribution from the contact regions is
minimized, and an accuracy of  0.5\,\textperthousand\- of a pH shift in one Hz bandwidth can be reached.
\end{abstract}

\maketitle

During the past decade, there has been a growing interest in applying the concept of an ion-sensitive field effect
transistor~\cite{Bergveld70,Bergveld03} to nano\-scale devices. It has been shown that
carbon nano\-tube,\cite{Tans98,Kong00,Collins00,Krueger01} graphene,\cite{Ang08,Ohno09,Cheng10}
nano\-wire,\cite{Duan01,Cui01,Stern07} and nano\-ribbon (NR) FETs~\cite{Nikolaides04,Elfstroem08}
are especially promising for sensing applications. Compared to conventional FETs,
nano\-scaled devices provide a larger surface-to-volume ratio. This results in a high sensitivity of the overall FET channel
conductance to changes in the surface potential caused by the adsorption of molecules.\cite{Elfstrom07} In order to reach
the detection limit, intense attempts have recently been made to understand the factors determining the signal-to-noise
ratio (SNR).\cite{Heller09,Gao10,Cheng10,Rajan10} Studies on carbon nano\-tube FETs~\cite{Heller09} and on nano\-wire FETs~\cite{Gao10}
showed that the SNR increases in the subthreshold regime, which is therefore the preferred regime for high sensitivity. However, a more
detailed understanding of the noise properties is needed to optimize the SNR across the full operating range of the FET.

In the present work, we measure the low-frequency $1/f$ noise of a dual-gated NR-FET~\cite{Elibol08, Heller09, Knopfmacher09,Knopfmacher10}
in ambient and in a buffer solution and determine the resolution limit expressed in a noise equivalent
threshold voltage shift $\delta V_{th}$, the latter being the measurement quantity in these types of sensors. We identify
two regimes which differ in the relative importance of the contact and intrinsic NR resistance. The lowest
value in $\delta V_{th}$ is found when the working point of the NR-FET is adjusted such that the
intrinsic NR resistance dominates. In the other case when the contact resistance dominates the noise can be larger by almost
two orders of magnitude for nominally the same overall resistance. This result shows the importance of being able to adjust
the operating point properly. In the best possible case we determine a resolution limit of \mbox{0.5\,\textperthousand} of a
pH change in one Hz bandwidth, which is comparable to a commercial pH meter (0.1\,\%~\cite{Microsens}), but for
a much smaller active sensing area.

\begin{figure}
\includegraphics[width=85mm]{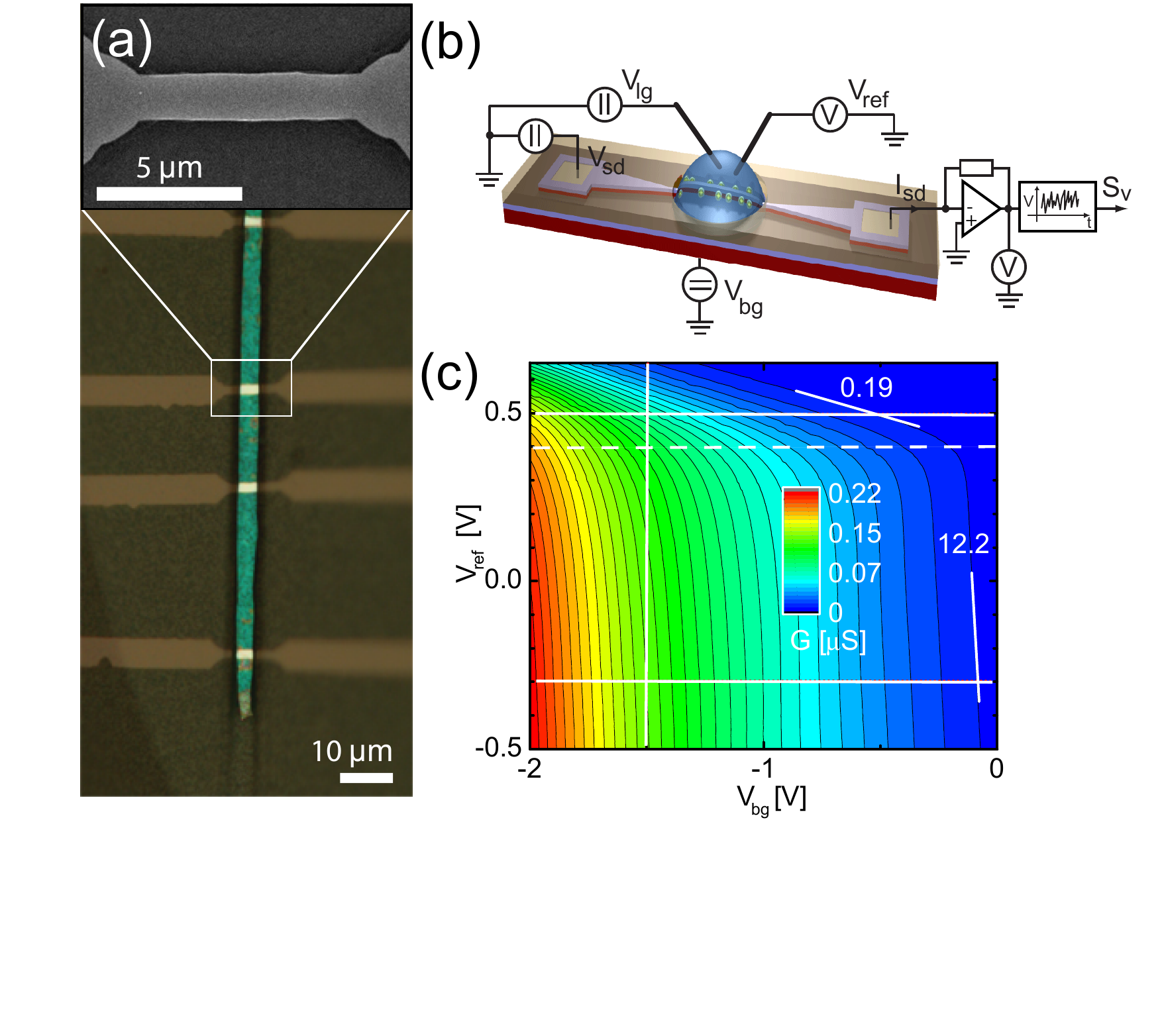}
  \caption{\label{fig1}(a) Optical image of a sample with four nano\-ribbons (NRs) (horizontal) and an enlarged SEM image of one of the NRs.
  The (vertical) liquid channel is the only part of the sample which is not covered with photo\-resist.
  (b) A schematic representation of the setup used for the measurements in liquid. There are two gates, a back-gate
  and a liquid-gate with applied gate voltages $V_{bg}$ and $V_{lg}$. The liquid potential is measured by a calomel reference
  electrode and denoted as $V_{ref}$. The NR-FETs are characterized by their small-signal conductance map $G(V_{bg},V_{ref})$, shown
  in (c), and by the noise obtained from the temporal dependence of the source-drain current $I_{sd}$ using fast Fourier transform.
  In (c) the horizontal dashed line marks the border between two regimes, the NR (upper) and contact (lower) dominated regime.
  Noise measurements were conducted along the three solid white lines. Short solid lines and numbers represent
  the slopes of the equi\-conductance lines.}
\end{figure}

%
Silicon NR-FETs were produced by UV lithography~\cite{Knopfmacher10} according to the top-down approach introduced
by Nikolaides {\it et al.}~\cite{Nikolaides04} and further developed to nano\-wire FETs by Stern {\it et al.}~\cite{Stern07}.
This high-yield process provides reproducible, hysteresis-free FETs with the following
dimensions: length\,$\times$\,width\,$\times$\,height = 10\,$\mu$m\,$\times$\, 700\,nm\,$\times$\,80\,nm (Fig.~\ref{fig1}a).
A thin Al$_2$O$_3$ layer was deposited on the device to ensure leakage-free operation in an electrolyte solution.
In addition, a liquid channel was formed in a photo\-resist layer, reducing the total area exposed to the electrolyte.

The measurement setup is schematically shown in Fig.~\ref{fig1}b. The NR FETs were operated at low source-drain DC
voltages $V_{sd}=10-100$\,mV in the linear regime. The source-drain current $I_{sd}$ through the NR was measured by a
current-voltage converter with a variable gain ($10^5-10^9$\,V/A). The conductance $G$ of the NR-FET is then obtained
as the ratio $G=I_{sd}/V_{sd}$ while varying both the back-gate $V_{bg}$ and liquid gate voltage $V_{lg}$.
This yields a two-dimensional (2D) conductance map, as shown in Fig.~\ref{fig1}c~\cite{Knopfmacher10}.
The vertical axis $V_{ref}$ is the potential of the liquid, as measured by a calomel reference electrode.
The equivalent voltage noise power spectral density $S_{V}$ (see e.\,g.~\cite{Hooge81})
was determined along the solid white lines through a fast Fourier transform of the time dependent fluctuations
of $I_{sd}$.

The conductance map in Fig.~\ref{fig1}c  displays two different regimes, above and below the white dashed line
at about $V_{ref}=+0.4$\,V, which differ in their relative coupling to the two gates. This is visible in the slopes
$s=\partial V_{ref}/\partial V_{bg}$ (short white lines and numbers), determined at constant $G$, that represent the
ratio of the gate coupling capacitances $C_{bg}/C_{lg}$, where $C_{bg}$ denotes the back-gate and $C_{lg}$
the liquid-gate capacitance~\cite{Knopfmacher10}. To understand the origin of the two different regimes,
one has to see that the NR-FET resistance $R=1/G$ is composed of two resistances in series. The intrinsic resistance
$R_{NR}$ and the contact resistance $R_{c}$. Due to the confinement of the liquid channel to the NR, $R_c$ is only weakly affected by
the liquid gate (small $C_{lg})$. Hence, if $R_c$ dominates $R$, $C_{bg}/C_{lg}$ is large, which corresponds to the lower regime
with $V_{ref} < 0.4$\,V. In contrast, if $R_c$ can be neglected, $R$ is determined by $R_{NR}$, which on its own is more strongly
capacitively coupled to the liquid than to the back-gate. We refer to the two regimes as contact and NR dominated.

\begin{figure}
\includegraphics[width=85mm]{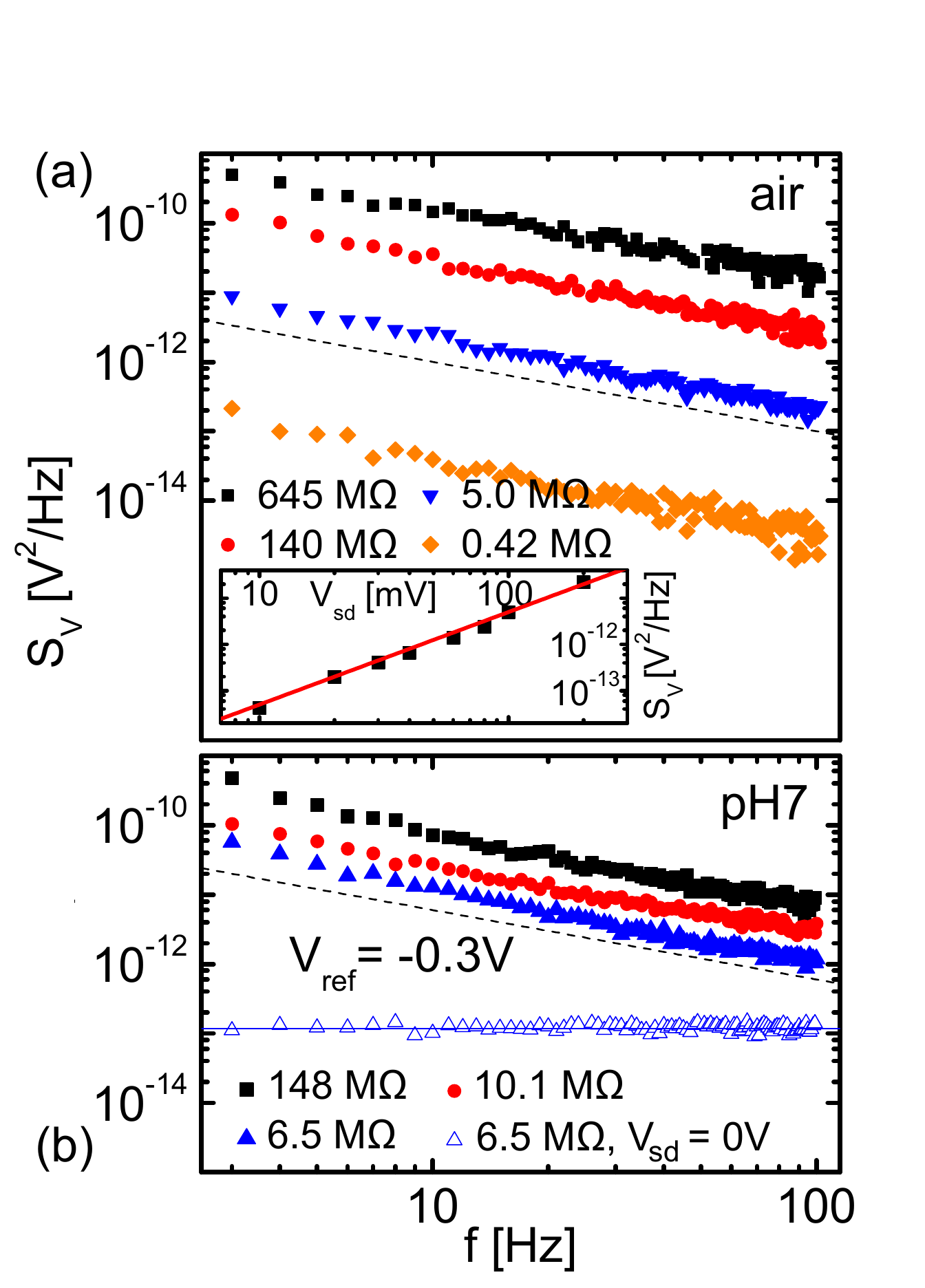}
  \caption{\label{fig2}
  The noise power spectral density of the voltage fluctuations $S_V$ obtained for a source-drain bias voltage of $V_{sd}=90$\,mV
  for different resistances of a NR measured (a) in air and (b) in a buffer solution (Titrisol pH 7, Merck).
  The dashed lines indicate a $1/f$ slope. The open symbols in (b) represent the thermal noise of the NR at $V_{sd}=0$\,V
  (\textcolor{blue}{$\triangle$}). The calculated thermal noise of a 6.5\,M$\Omega$
  resistor at 300\,K is shown for comparison (horizontal line).
  Inset: $S_V$ as a function of $V_{sd}$ at $7$\,Hz, measured in air (logarithmic scale).
  The solid red line indicates a power law with exponent two.}
\end{figure}

Figure~\ref{fig2} shows the frequency dependence of the noise power $S_V(f)$ of the NR for different resistance values,
measured in air (a) and in a buffer solution with pH 7 (b). The corresponding thermal background noise, recorded at zero bias,
has been subtracted from the data. An example is shown by (\textcolor{blue}{$\triangle$}). $S_V(f)$ has a clear $1/f$ dependence (dashed lines),
and its amplitude is proportional to $V_{sd}^2$ (inset), as expected for $1/f$ noise~\cite{Hooge81}.
Such a behavior can phenomenologically be described by Hooge's law~\cite{Hooge69, Hooge81}
\begin{equation}
  \label{Hooge}
  S_V(f)=V_{sd}^2\frac{\alpha}{Nf}.
\end{equation}
The material dependent parameter $\alpha$ accounts for scattering effects and the constant $N$ denotes the number of
fluctuators in the system.

\begin{figure}
\includegraphics[width=85mm]{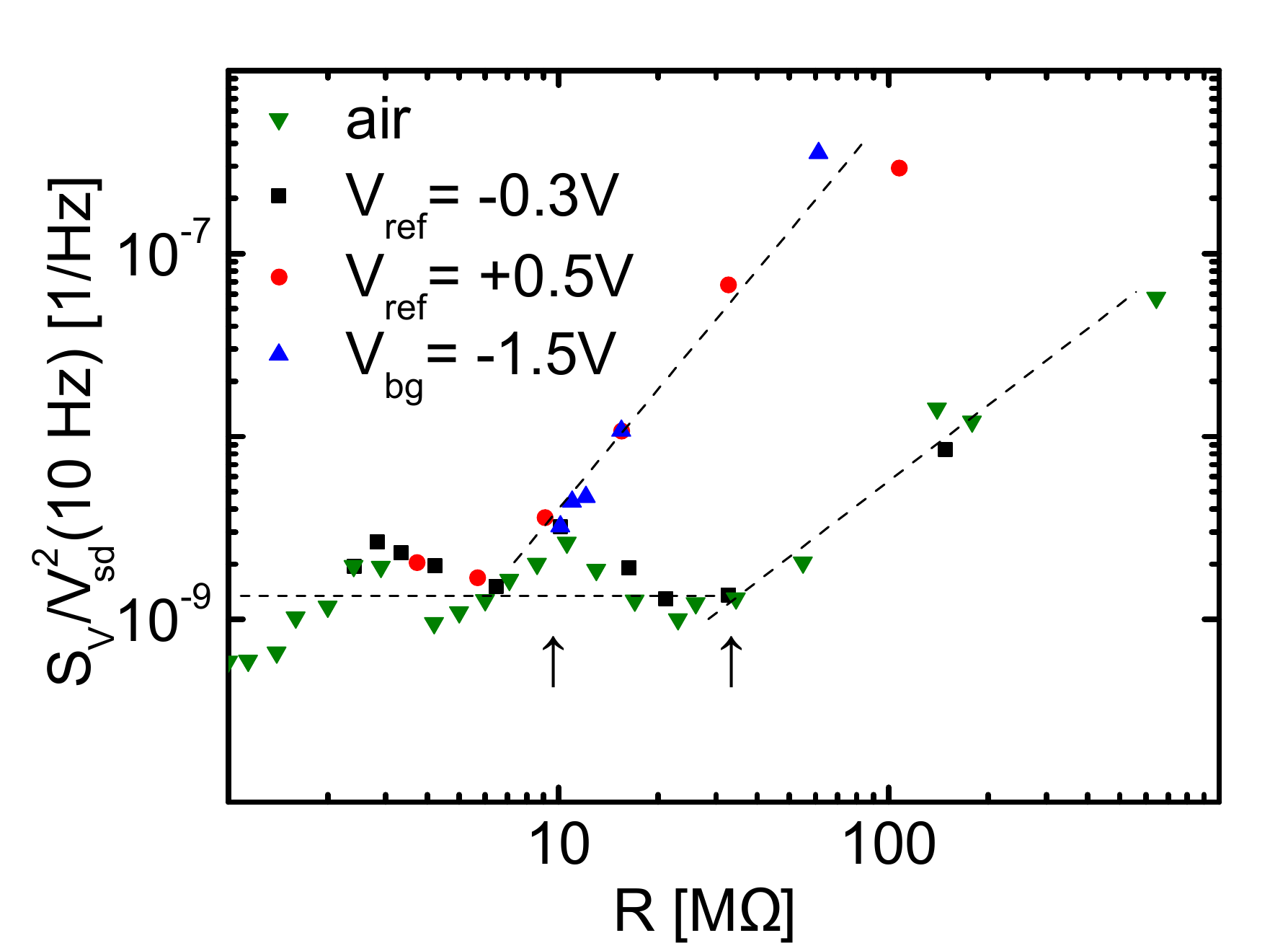}
  \caption{\label{fig3} $S_V$ divided by the squared source-drain voltage $V_{sd}^2$
  as a function of $R$ at 10\,Hz, measured in air and in buffer solution (logarithmic scale).
  Dashed lines are guides to the eye. Arrows indicate the transition between different regimes.}
\end{figure}

In Fig.~\ref{fig3}, the normalized noise amplitude $S_V/V_{sd}^2$ at 10\,Hz is depicted as a function of $R$. The noise in
the system increases dramatically (indicated by dashed lines) above a certain threshold resistance value. The position of this
threshold (arrows) and the steepness of the rise depends on whether the NR is gated by the liquid or the back-gate.
In air ($\blacktriangledown$), where $V_{bg}$ is the only applied gate voltage, the noise starts to increase at roughly 30\,M$\Omega$.
A similar behavior is observed in liquid in the contact-dominated regime, i.e. for $V_{ref}=-0.3$\,V ({\small $\blacksquare$}).
In contrast, within the NR-dominated regime ($\bullet$),
the noise increases steeper, starting already at about 10\,M$\Omega$. For $R$ smaller than the respective thresholds,
the noise level is approximately constant. The apparent superimposed structure observed in this range is wire specific.
Different NR-FETs, while confirming the general dependence, typically display a different fine structure. The thresholds of $10-30$ M$\Omega$ correspond to the transition from the linear to
the subthreshold regime of the FETs.

The physical signal in FET sensors is the shift $\Delta V_{th}$ of the threshold voltage $V_{th}$ caused by a chemical change on the sensing surface.
It is obtained from the measured conductance change $\Delta G$ and the trans\-conductance $g=G'=dG(V_g)/dV_g$, characteristic for a given FET, as
$\Delta V_{th}=\Delta G/g$. This equation can be used to determine the true figure of merit which is the equivalent noise power
of the threshold voltage $\delta V_{th}$ given by
\begin{equation}
  \label{deltaVth}
  \delta V_{th}=\frac{\sqrt{S_V/V_{sd}^2(f)}}{g/G}=\frac{\sqrt{S_V/V_{sd}^2(f)}}{(\ln G)'}.
\end{equation}
Here, we have made use of the relation $\delta G/G=\sqrt{S_V}/V_{sd}$.

\begin{figure}
\includegraphics[width=85mm]{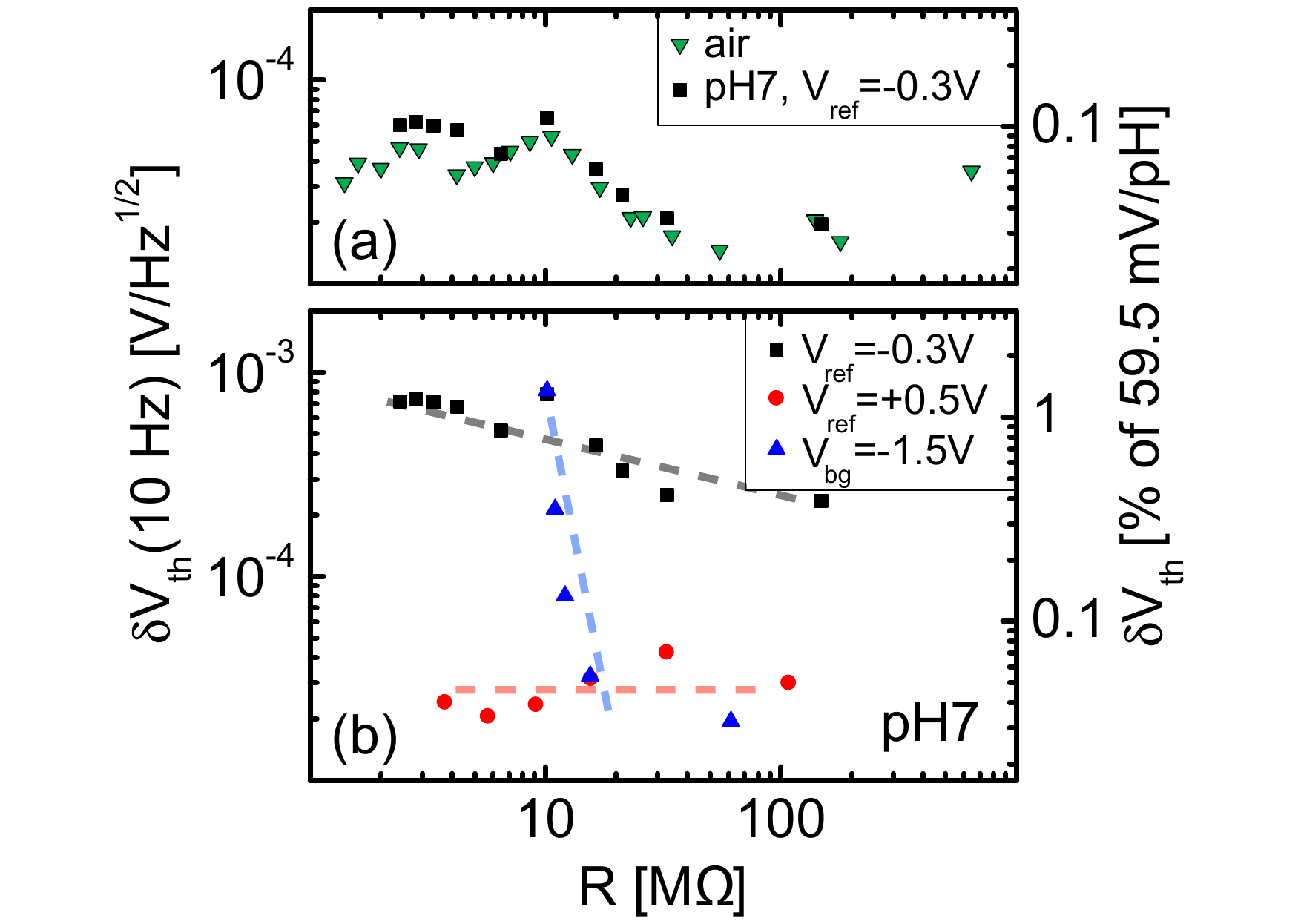}
  \caption{\label{fig4} Threshold voltage fluctuations $\delta V_{th}$=$S_V/(\ln G(V_{bg}))'$
  (a) and $S_V/(\ln G(V_{ref}))'$ (b), calculated from the data in Fig. \ref{fig3} as a function of $R$.
  Dashed lines are guides to the eye. The right axis shows $\delta V_{th}$ relative to the Nernst limit of the
  pH sensitivity (59.5\,mV/pH at 300\,K).}
\end{figure}

In Fig.~\ref{fig4}a we show $\delta V_{th}$ when the controlling gate is $V_{bg}$
for data measured in air (\textcolor{darkgreen}{$\blacktriangledown$}) together with
the data acquired in buffer solution at $V_{ref}=-0.3$\,V ({\small $\blacksquare$}). Both curves show a very similar behavior.
Since we know that the liquid data obtained at $V_{ref}=-0.3$\,V is contact dominated, we conclude that the measurement in air is also
contact dominated.

In Fig.~\ref{fig4}b we summarize $\delta V_{th}$ for measurements done in an electrolyte. To obtain $\delta V_{th}$, we consistently use
$V_{lg}$ as the controlling gate for all three data sets in this figure. Interestingly, in the NR-dominated regime ($\bullet$) $\delta V_{th}$
is much smaller than in the contact-dominated regime ({\small $\blacksquare$}). The difference can amount to almost two orders of magnitude.
Although the voltage noise values $S_V$ are not much different in the two regimes (Fig.~\ref{fig3}), the sensitivities in the true measurement
quantity greatly differ. This shows that the trans\-conductance values, and therefore the gate-coupling to the liquid, are crucial
factors determining the ultimate sensitivity. We also stress that $\delta V_{th}$ can be low over an extended range of NR resistance values
$R$, from $\sim 1$ to $100$\,M$\Omega$. This range covers the transition from the linear to the subthreshold regime.
The lowest value of $2-3\cdot 10^{-5}$\,V$/\sqrt{\textnormal{Hz}}$ corresponds to an accuracy of
0.5\,\textperthousand\- of a typical Nernstian pH shift in one Hz bandwidth (right axis)
throughout the full resistance range ({$\bullet$}).

The data set obtained at a fixed $V_{bg}$ and varying $V_{ref}$ (\textcolor{blue}{$\blacktriangle$}) demonstrate the cross-over between the two
different regimes. In this case a very pronounced transition from a regime with low sensitivity (low $R$) to a regime with high sensitivity
(larger $R$) is apparent. For this case, it has recently been pointed out~\cite{Gao10} that the signal-to-noise ratio (SNR)
(corresponding to $1/\delta V_{th}$) increases with resistance $R$ and is the highest in the subthreshold regime.
We confirm this as well but we emphasize that the dual-gate approach used here provides a more general and detailed insight.
For $V_{ref}=-0.3$\,V, the contact leads also contribute to the total noise and strongly decrease the SNR.
In contrast, for $V_{ref}=+0.5$\,V, the resistance of the NR-FET is not contact-dominated. In that case the SNR is constantly large over
the whole resistance range.

As a last step, we estimate the charge noise of the NR, which corresponds to the minimum detectable number of charge carriers on the gate.
To do so, we first define the gate-related power spectrum voltage noise $S_{Vg}=S_I/g_m^2$. Here, $S_I=S_V/R^2$ is the current noise
that can be easily determined from Fig.~\ref{fig3}, and $g_m=\partial I/\partial V_g$ denotes the trans\-conductance with respect
to the controlling gate. For the measurement versus the liquid gate ($V_{bg}=-1.5$\,V), we obtain a charge noise at 10\,Hz of
$\sqrt{S_q}=C_{lg}\cdot \sqrt{S_{Vg}}/e\approx 5.8$\,$e/\sqrt{\textnormal{Hz}}$, where $C_{lg}\approx 26$\,fF is the estimated liquid
gate capacitance in our system and $e$ is the elementary charge.



In conclusion, we have studied the low-frequency noise in dual-gated Si nano\-ribbon-FET sensors and
determined the signal-to-noise ratio over a large resistance range. The deduced threshold voltage noise $\delta V_{th}$ is an important
quantity in a FET sensor and strongly depends on the working point. We stress that $\delta V_{th}$ can be low over an extended
range from the linear to the subthreshold regime, even though the voltage noise $S_V$ grows non-linearly with
resistance and is the highest in the subthreshold range. We also confirmed recent studies that found the SNR
increasing with resistance in a certain case.

\begin{acknowledgments}
The authors acknowledge the LMN at PSI Villigen for the oxidation of the SOI wafers.
We are grateful for the support provided by the nano-tera.ch, Sensirion AG, and the Swiss Nanoscience Institute (SNI).
\end{acknowledgments}


\end{document}